\documentclass[10pt]{article}
\usepackage[OE]{express}
\usepackage{mathtools}
\usepackage{algorithm}
\usepackage[noend]{algpseudocode}
\makeatletter
\def\BState{\State\hskip-\ALG@thistlm}
\makeatother
\DeclarePairedDelimiter{\ceil}{\lceil}{\rceil}
\usepackage{enumitem}

\begin{document}
\title{Signal Processing Based Pile-up Compensation for Gated Single-Photon Avalanche Diodes}

\author{Adithya Kumar Pediredla,\authormark{1,*} Aswin C. Sankaranarayanan,\authormark{2} Mauro Buttafava,\authormark{3} Alberto Tosi,\authormark{3} and Ashok Veeraraghavan\authormark{1}}

\address{\authormark{1}Electrical and Computer Engineering, Rice University, 6100 Main street, Houston, Texas 77005, USA\\
\authormark{2} Electrical and Computer Engineering, Carnegie Mellon University, 5000 Forbes Ave, Pittsburgh 15213, USA\\
\authormark{3}Dipartimento di Elettronica, Informazione e Bioingegneria, Politecnico di Milano, Milan, Italy}

\email{\authormark{*}akp4@rice.edu} 

\begin{abstract}
	Single-photon avalanche diode (SPAD) based transient imaging suffers from an aberration called pile-up. 
	When multiple photons arrive within a single repetition period of the illuminating laser, the SPAD records only the arrival of the first photon; this leads to a bias in the recorded light transient wherein the transient response at later time-instants are under-estimated. 
	An unfortunate consequence of this is the need to operate the illumination at low-power levels to reduce the probability of multiple photons returning in a single period. 
	Operating the laser at low power results in either low signal-to-noise ratio (SNR) in the measured transients or reduced frame rate due to longer exposure durations to achieve a high SNR.
	In this paper, we propose a signal processing-based approach to compensate pile-up in post-processing, thereby enabling high power operation of the illuminating laser.
	While increasing illumination does cause a  fundamental information loss in the data captured by SPAD, we quantify this information loss using Cramer-Rao bound and show that the errors in our framework are only limited to this information loss.
	We experimentally validate our hypotheses using real data from a lab prototype.
\end{abstract}
\ocis{(000.5490) Probability theory, stochastic processes, and statistics; (010.3640) Lidar; (040.1345) Avalanche photodiodes (APDs); (320.0320) Ultrafast optics;}

\bibliographystyle{osajnl}
\bibliography{SPSPAD}

\begin{thebibliography}{10}
\newcommand{\enquote}[1]{``#1''}

\bibitem{jarabo2017recent}
A.~Jarabo, B.~Masia, J.~Marco, and D.~Gutierrez, \enquote{Recent advances in
  transient imaging: A computer graphics and vision perspective,} Visual
  Informatics \textbf{1}, 65--79 (2017).

\bibitem{velten2013femto}
A.~Velten, D.~Wu, A.~Jarabo, B.~Masia, C.~Barsi, C.~Joshi, E.~Lawson,
  M.~Bawendi, D.~Gutierrez, and R.~Raskar, \enquote{Femto-photography:
  capturing and visualizing the propagation of light,} ACM Transactions on
  Graphics (ToG) \textbf{32}, 44 (2013).

\bibitem{ICCDCamera}
\enquote{{Ultra high speed ICCD camera},}
  \url{http://stanfordcomputeroptics.com/download/Brochure-4Picos.pdf}.
  Accessed: 2018-02-19.

\bibitem{gariepy2015single}
G.~Gariepy, N.~Krstaji{\'c}, R.~Henderson, C.~Li, R.~R. Thomson, G.~S. Buller,
  B.~Heshmat, R.~Raskar, J.~Leach, and D.~Faccio, \enquote{Single-photon
  sensitive light-in-fight imaging,} Nature Communications \textbf{6} (2015).

\bibitem{o2017reconstructing}
M.~O'Toole, F.~Heide, D.~B. Lindell, K.~Zang, S.~Diamond, and G.~Wetzstein,
  \enquote{Reconstructing transient images from single-photon sensors,} in
  \enquote{Proc. IEEE CVPR,}  (2017), pp. 2289--2297.

\bibitem{franch2017low}
N.~Franch, O.~Alonso, J.~Canals, A.~Vil{\`a}, and A.~Dieguez, \enquote{A low
  cost fluorescence lifetime measurement system based on spad detectors and
  fpga processing,} Journal of Instrumentation \textbf{12}, C02070 (2017).

\bibitem{shin2015photon}
D.~Shin, A.~Kirmani, V.~K. Goyal, and J.~H. Shapiro, \enquote{Photon-efficient
  computational 3-d and reflectivity imaging with single-photon detectors,}
  IEEE Transactions on Computational Imaging \textbf{1}, 112--125 (2015).

\bibitem{satat2017object}
G.~Satat, M.~Tancik, O.~Gupta, B.~Heshmat, and R.~Raskar, \enquote{Object
  classification through scattering media with deep learning on time resolved
  measurement,} Optics express \textbf{25}, 17466--17479 (2017).

\bibitem{pediredla2017linear}
A.~K. Pediredla, N.~Matsuda, O.~Cossairt, and A.~Veeraraghavan, \enquote{Linear
  systems approach to identifying performance bounds in indirect imaging,} IEEE
  International Conference on Acoustics, Speech and Signal Processing  (2017).

\bibitem{pifferi2008time}
A.~Pifferi, A.~Torricelli, L.~Spinelli, D.~Contini, R.~Cubeddu, F.~Martelli,
  G.~Zaccanti, A.~Tosi, A.~Dalla~Mora, F.~Zappa, and S.~Cova,
  \enquote{Time-resolved diffuse reflectance using small source-detector
  separation and fast single-photon gating,} Physical Review Letters
  \textbf{100}, 138101 (2008).

\bibitem{bollinger1961measurement}
L.~Bollinger and G.~E. Thomas, \enquote{Measurement of the time dependence of
  scintillation intensity by a delayed-coincidence method,} Review of
  Scientific Instruments \textbf{32}, 1044--1050 (1961).

\bibitem{becker2005advanced}
W.~Becker, \emph{Advanced time-correlated single photon counting techniques},
  vol.~81 (Springer Science \& Business Media, 2005).

\bibitem{becker2015advanced}
W.~Becker, \emph{Advanced time-correlated single photon counting applications}
  (Springer, 2015).

\bibitem{o2012time}
D.~O'Connor, \emph{Time-correlated single photon counting} (Academic Press,
  2012).

\bibitem{cova1981towards}
S.~Cova, A.~Longoni, and A.~Andreoni, \enquote{Towards picosecond resolution
  with single-photon avalanche diodes,} Review of Scientific Instruments
  \textbf{52}, 408--412 (1981).

\bibitem{schwartz2008single}
D.~E. Schwartz, E.~Charbon, and K.~L. Shepard, \enquote{A single-photon
  avalanche diode array for fluorescence lifetime imaging microscopy,} IEEE
  journal of solid-state circuits \textbf{43}, 2546--2557 (2008).

\bibitem{panzeri2013single}
F.~Panzeri, A.~Ingargiola, R.~R. Lin, N.~Sarkhosh, A.~Gulinatti, I.~Rech,
  M.~Ghioni, S.~Cova, S.~Weiss, and X.~Michalet, \enquote{Single-molecule fret
  experiments with a red-enhanced custom technology spad,} Single Molecule
  Spectroscopy and Superresolution Imaging VI \textbf{8590}, 85900D (2013).

\bibitem{li1993single}
L.-Q. Li and L.~M. Davis, \enquote{Single photon avalanche diode for single
  molecule detection,} Review of Scientific Instruments \textbf{64}, 1524--1529
  (1993).

\bibitem{gyongy2016smart}
I.~Gyongy, A.~Davies, N.~A. Dutton, R.~R. Duncan, C.~Rickman, R.~K. Henderson,
  and P.~A. Dalgarno, \enquote{Smart-aggregation imaging for single molecule
  localisation with spad cameras,} Scientific Reports \textbf{6} (2016).

\bibitem{krishnaswami2014towards}
V.~Krishnaswami, C.~J. Van~Noorden, E.~M. Manders, and R.~A. Hoebe,
  \enquote{Towards digital photon counting cameras for single-molecule optical
  nanoscopy,} Optical Nanoscopy \textbf{3}, 1 (2014).

\bibitem{cheng2016cmos}
Z.~Cheng, \enquote{{CMOS}-based single photon avalanche diode and
  time-to-digital converter towards pet imaging applications,} Ph.D. thesis
  (2016).

\bibitem{torricelli2014time}
A.~Torricelli, D.~Contini, A.~Pifferi, M.~Caffini, R.~Re, L.~Zucchelli, and
  L.~Spinelli, \enquote{Time domain functional nirs imaging for human brain
  mapping,} Neuroimage \textbf{85}, 28--50 (2014).

\bibitem{mazurenka2012development}
M.~Mazurenka, H.~Wabnitz, A.~Dalla~Mora, D.~Contini, A.~Pifferi, R.~Cubeddu,
  A.~Tosi, F.~Zappa, and R.~Macdonald, \enquote{Development of an optical
  non-contact time-resolved diffuse reflectance scanning imaging system,}
  Biomedical Optics pp. BTu3A--50 (2012).

\bibitem{mazurenka2012non}
M.~Mazurenka, A.~Jelzow, H.~Wabnitz, D.~Contini, L.~Spinelli, A.~Pifferi,
  R.~Cubeddu, A.~Dalla~Mora, A.~Tosi, F.~Zappa \emph{et~al.},
  \enquote{Non-contact time-resolved diffuse reflectance imaging at null
  source-detector separation,} Optics Express \textbf{20}, 283--290 (2012).

\bibitem{di2016characterization}
L.~Di~Sieno, H.~Wabnitz, A.~Pifferi, M.~Mazurenka, Y.~Hoshi, A.~Dalla~Mora,
  D.~Contini, G.~Boso, W.~Becker, F.~Martelli \emph{et~al.},
  \enquote{Characterization of a time-resolved non-contact scanning diffuse
  optical imaging system exploiting fast-gated single-photon avalanche diode
  detection,} Review of Scientific Instruments \textbf{87}, 035118 (2016).

\bibitem{maccarone2015underwater}
A.~Maccarone, A.~McCarthy, X.~Ren, R.~E. Warburton, A.~M. Wallace, J.~Moffat,
  Y.~Petillot, and G.~S. Buller, \enquote{Underwater depth imaging using
  time-correlated single-photon counting,} Optics Express \textbf{23},
  33911--33926 (2015).

\bibitem{niclass20073d}
C.~Niclass, M.~Soga, and E.~Charbon, \enquote{3d imaging based on single photon
  detectors,} 2nd Symposium on Range Imaging  (2007).

\bibitem{kirmani2014first}
A.~Kirmani, D.~Venkatraman, D.~Shin, A.~Cola{\c{c}}o, F.~N. Wong, J.~H.
  Shapiro, and V.~K. Goyal, \enquote{First-photon imaging,} Science
  \textbf{343}, 58--61 (2014).

\bibitem{shin2016photon}
D.~Shin, F.~Xu, D.~Venkatraman, R.~Lussana, F.~Villa, F.~Zappa, V.~K. Goyal,
  F.~N. Wong, and J.~H. Shapiro, \enquote{Photon-efficient imaging with a
  single-photon camera,} Nature Communications \textbf{7} (2016).

\bibitem{shin2016computational}
D.~Shin, J.~H. Shapiro, and V.~K. Goyal, \enquote{Computational single-photon
  depth imaging without transverse regularization,} IEEE International
  Conference on Image Processing  (2016).

\bibitem{buttafava2015non}
M.~Buttafava, J.~Zeman, A.~Tosi, K.~Eliceiri, and A.~Velten,
  \enquote{Non-line-of-sight imaging using a time-gated single photon avalanche
  diode,} Optics Express \textbf{23}, 20997--21011 (2015).

\bibitem{tsai2017geometry}
C.-Y. Tsai, K.~N. Kutulakos, S.~G. Narasimhan, and A.~C. Sankaranarayanan,
  \enquote{The geometry of first-returning photons for non-line-of-sight
  imaging,} CVPR  (2017).

\bibitem{gariepy2015detection}
G.~Gariepy, F.~Tonolini, R.~Henderson, J.~Leach, and D.~Faccio,
  \enquote{Detection and tracking of moving objects hidden from view,} Nature
  Photonics  (2015).

\bibitem{pediredla2017reconstructing}
A.~K. Pediredla, M.~Buttafava, A.~Tosi, O.~Cossairt, and A.~Veeraraghavan,
  \enquote{Reconstructing rooms using photon echoes: A plane based model and
  reconstruction algorithm for looking around the corner,} IEEE International
  Conference on Computational Photography  (2017).

\bibitem{heide2017robust}
F.~Heide, M.~O'Toole, K.~Zhang, D.~Lindell, S.~Diamond, and G.~Wetzstein,
  \enquote{Robust non-line-of-sight imaging with single photon detectors,}
  arXiv preprint arXiv:1711.07134  (2017).

\bibitem{coates1968correction}
P.~Coates, \enquote{The correction for photon pile-up' in the measurement of
  radiative lifetimes,} Journal of Physics E: Scientific Instruments
  \textbf{1}, 878 (1968).

\bibitem{coates1992analytical}
P.~Coates, \enquote{Analytical corrections for dead time effects in the
  measurement of time-interval distributions,} Review of Scientific Instruments
  \textbf{63}, 2084--2088 (1992).

\bibitem{davis1970single}
C.~Davis and T.~King, \enquote{Single photon counting pileup corrections for
  time-varying light sources,} Review of Scientific Instruments \textbf{41},
  407--408 (1970).

\bibitem{luhmann1997statistics}
T.~Luhmann, \enquote{Statistics and dead time correction of two-particle
  time-of-flight coincidence experiments,} Review of Scientific Instruments
  \textbf{68}, 2347--2356 (1997).

\bibitem{renfro1999pulse}
M.~Renfro, S.~Pack, G.~King, and N.~Laurendeau, \enquote{A pulse-pileup
  correction procedure for rapid measurements of hydroxyl concentrations using
  picosecond time-resolved laser-induced fluorescence,} Applied Physics B:
  Lasers and Optics \textbf{69}, 137--146 (1999).

\bibitem{walker2002iterative}
J.~G. Walker, \enquote{Iterative correction for pile-up' in single-photon
  lifetime measurement,} Optics Communications \textbf{201}, 271--277 (2002).

\bibitem{patting2007dead}
M.~Patting, M.~Wahl, P.~Kapusta, and R.~Erdmann, \enquote{Dead-time effects in
  {TCSPC} data analysis,} International Congress on Optics and Optoelectronics
  \textbf{6583}, 658307 (2007).

\bibitem{rebafka2011information}
T.~Rebafka, F.~Roueff, and A.~Souloumiac, \enquote{Information bounds and mcmc
  parameter estimation for the pile-up model,} Journal of Statistical Planning
  and Inference \textbf{141}, 1--16 (2011).

\bibitem{arlt2013study}
J.~Arlt, D.~Tyndall, B.~R. Rae, D.~D.-U. Li, J.~A. Richardson, and R.~K.
  Henderson, \enquote{A study of pile-up in integrated time-correlated single
  photon counting systems,} Review of Scientific Instruments \textbf{84},
  103105 (2013).

\bibitem{kay1993statistical}
S.~M. Kay, \enquote{Statistical signal processing,} Estimation Theory
  \textbf{1} (1993).

\bibitem{scott2015multivariate}
D.~W. Scott, \emph{Multivariate density estimation: theory, practice, and
  visualization} (John Wiley \& Sons, 2015).

\bibitem{DropboxLink}
\enquote{{Dataset captured and all the codes used in this paper},}
  \url{https://www.dropbox.com/sh/8zyk5r0qla4c6xc/AADns1HgBi40P1pU1YJwKie3a?dl=0}.
  Accessed: 2018-02-19.

\bibitem{cova1996avalanche}
S.~Cova, M.~Ghioni, A.~Lacaita, C.~Samori, and F.~Zappa, \enquote{Avalanche
  photodiodes and quenching circuits for single-photon detection,} Applied
  optics \textbf{35}, 1956--1976 (1996).

\bibitem{hernandez2017computational}
Q.~Hernandez, D.~Gutierrez, and A.~Jarabo, \enquote{A computational model of a
  single-photon avalanche diode sensor for transient imaging,} arXiv preprint
  arXiv:1703.02635  (2017).

\bibitem{ross2014introduction}
S.~M. Ross, \emph{Introduction to probability models} (Academic press, 2014).

\bibitem{buttafava2014time}
M.~Buttafava, G.~Boso, A.~Ruggeri, A.~Dalla~Mora, and A.~Tosi,
  \enquote{Time-gated single-photon detection module with 110 ps transition
  time and up to 80 {MHz} repetition rate,} Review of Scientific Instruments
  \textbf{85}, 083114 (2014).

\end{thebibliography}

\section{Introduction}

\begin{figure}
	\centering
	\includegraphics[width=5.250in]{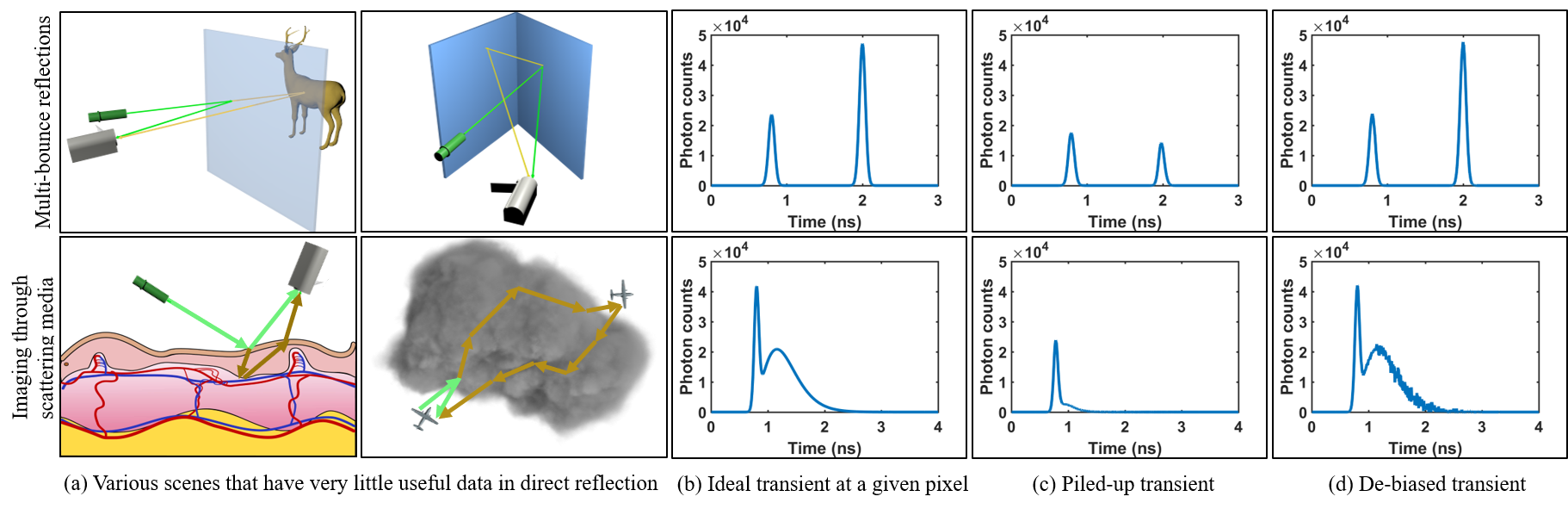} \\
	\caption{Transient response of a scene is the number of photons received by the camera as a function of time (typically sub-nanosecond resolution).  
		(a) For imaging through scattering media or imaging in the presence of multi-bounce reflections, the direct reflection carries little useful information. 
		(b) Recording the entire transient, however, will allow us to reconstruct the obscured objects. The SPAD-based TCSPC is one such device that can record transients if the illumination and the background intensities are low. 
		(c) If either the illumination or the background is high, the transients measured by the SPAD-based TCSPC suffer from pile-up distortion, which is a non-linear bias in the recorded transient. Pile-up distortion draws its name from that fact that most photons pile-up at the beginning of the transient. 
		(d) We propose a signal processing approach to remove the bias of piled-up transient measured.} 
	\label{Fig:MainFigure}
\end{figure}

Imaging at ultra-high speed is pivotal in many scientific disciplines. From demystifying the hovering of a hummingbird to the accurate hunts of dragonflies, increasing the speed of imaging has unlocked many secrets in nature. 
%
Today, high-speed imaging is capable of imaging at pico-second time scales \cite{jarabo2017recent}; at such resolutions, we can observe light-in-flight or light as it propagates in a scene \cite{velten2013femto,ICCDCamera,gariepy2015single,o2017reconstructing}.
One approach for achieving this is by using a single-photon avalanche diode (SPAD) coupled with time-correlated single photon counting (TCSPC) electronics; this device has found use in fluorescence-life-time-imaging\ \cite{franch2017low}, depth sensing\ \cite{shin2015photon}, imaging through scattering medium\ \cite{satat2017object}, and imaging beyond line-of-sight\ \cite{gariepy2015single, pediredla2017linear} (Figure~\ref{Fig:MainFigure}). 

SPADs work by detecting the first-arriving photon after the scene is illuminated by a laser pulse. The illumination source, typically sends few millions picosecond pulses per second. 
By binning the time of arrival of the first-arriving photons after each laser pulse, the TCSPC records a histogram of photon counts as a function of arrival time.
The measured histogram is  proportional to the transient response if the probability of measuring a photon in a time bin is proportional to the transient response.
Unfortunately, this is true only when the laser illumination power is sufficiently low so that there is at most one returning photon after each pulse of the laser.
To understand why, note that the probability of photon arrivals increases with laser power.
Suppose that the laser power is large enough so that multiple photons return in each cycle.
The SPAD  only records the first returning photon in each cycle and, hence, does not bin the later-arriving photons.
This causing a systematic bias between the measured histogram and the true transient response; in particular, this bias is such that there is an under-counting of photons with larger time of arrival.
This effect is called pile-up, since the photons appear to pile-up near the origin due to uneven bias introduced in the transient response (Figure~\ref{Fig:MainFigure}(c)). 
In the limit, when the laser power is extremely high, we will not measure any photons with large time of arrivals.
An undesirable consequence of pile-up is that SPAD-based TCSPC is often performed at very low illumination power which leads to inefficiencies in the form of long exposure times and poor SNR.

%

To avoid pile-up, Pifferi et al.\ \cite{pifferi2008time} propose a gated use of the SPADs where in the SPAD is selectively turned off for some duration  every cycle. 
Typically the gate is used to block photons from the high-photon rate parts of the transient such as to decrease their probability of photon arrival. The gated operation decreases the pile-up effect significantly, but the number of cycles the SPAD does not detect any photon remains high. Gated-mode SPAD operation is also ineffective in  the presence of strong background where-in the high photon-rate regions of the transient is not temporally localized.

In this paper, we propose a signal-processing approach that enables the operation of gated SPADs even in the presence of high illumination power; we show that the number of cycles the SPAD detects a photon can be increased to 90\% from the typical current efficiency of $5\%$. 
%
The transient recorded by the TCSPC circuit is still biased, but we show that it can be unbiased via the use of appropriate post-processing techniques (Figure~\ref{Fig:MainFigure}(d)).
Our approach works in tandem with the SPAD gate and further increases the intensity levels the SPAD can accurately record transients. 

\paragraph{Contributions:} We make the following contributions:
\begin{enumerate}[leftmargin=*]
	\item Estimation-theoretic framework for modeling pile-up: We derive maximum-likelihood (ML), and maximum a posteriori (MAP) estimates for the transient response of the SPAD. We also derive an asymptotic bound (Cramer-Rao bound) on the estimation errors. 
	\item Experimental validation: We have built experimental hardware with the SPAD and TCSPC devices and showed that the SNR of the transients improves with our estimation framework. We also show that even in the presence of strong ambient illumination that causes the pile-up, we can recover the underlying signal.  
\end{enumerate}
\paragraph{Limitations:} The signal processing approach provided in this paper compensates for the pile-up effect that happens due to the dead time of the SPAD and TCSPC devices. However, the SPADs also suffer from finite quantum efficiency, dark counts, after pulsing, gate-ringing, and time-walk artifacts. Our theory does not account for these effects. The quantum efficiency is a linear phenomenon equivalent to decreasing the illumination power by a constant amount and hence, can be neglected. Other non-linear effects are minor, as the specs of a typical SPAD (refer Section~\ref{sec:SetUp}) show that our approximations are reasonable.
\section{Prior work}
Time-correlated single photon counting was first introduced by Bollinger and Thomas \cite{bollinger1961measurement} in 1961. A lot of progress was made since then, including the works of Becker \cite{becker2005advanced,becker2015advanced}, O'Connor and Desmund \cite{o2012time}. Usage of SPADs for TCSPC for pico-second temporal resolution was first shown by Cova et al.\ \cite{cova1981towards} in 1981. Since then, a lot of novel applications at both microscopic scale and macroscopic scale were designed with the SPAD-based TCSPCs. 
Imaging applications that can image objects beyond the line-of-sight at macroscopic scale were also becoming famous. Below, we review some of the recent advances in microscopic and macroscopic imaging applications with the SPAD. 

\subsection{Microscopy and Spectroscopy}
Schwartz et al.\ \cite{schwartz2008single} designed and characterized a fully integrated SPAD imager with a time-to-digital converter (TDC) for fluorescence lifetime imaging.  
Panzeri et al.\ \cite{panzeri2013single} characterized the use of SPADs in single-molecule fluorescence resonant energy (FRET) studies on freely diffusing molecules in confocal and alternating laser excitation schemes. 
Li et al.\ \cite{li1993single} first proposed to detect single molecules with SPADs. 
Recently, SPADs are also employed in other single-molecule studies like localization \cite{gyongy2016smart} or optical nanoscopy \cite{krishnaswami2014towards}.
Cheng et al.\ \cite{cheng2016cmos} developed a non-invasive positron-emission tomography with SPADs. 
The SPAD-based TCSPCs were extensively used in time-domain functional near infrared spectroscopy (NIRS) imaging for human brain mapping \cite{torricelli2014time}.
Mazurenka et al.\ \cite{mazurenka2012development,mazurenka2012non} developed a non-contact time-resolved diffuse reflectance imaging using a fast gated single photon counting for detection of absorption changes few centimeters deep of tissue (turbid medium). Sieno et al. \cite{di2016characterization} characterized these non-contact diffuse optical imaging systems.

\subsection{Macroscopy}
Many of the macroscopic applications targeted low noise and high sensitivity of SPADs. For example, Maccarone et al.\ \cite{maccarone2015underwater} designed a SPAD-based low power underwater depth imaging system. 
Niclass et al.\ \cite{niclass20073d} proposed a fast 3D imaging sensor with SPADs based on the time of arrival of the photons. 
Kirmani et al.\ \cite{kirmani2014first} proposed a low-photon-flux imaging system that can reconstruct both depth and reflectivity of a scene from the first arriving photons. 
Shin et al.\ \cite{shin2016photon,shin2016computational} extended this system that can reproduce similar results even with a nano-second jitter system but with prior knowledge on the scene. 
Gariepy et al.\ \cite{gariepy2015single} used a 32$\times$32 SPAD array to capture transient images directly at 67 ps temporal resolution. 
Buttafava et al.\ \cite{buttafava2015non} repurposed SPADs to look around the corners, and reconstruct objects beyond line-of-sight. 
Pediredla et al.\ \cite{pediredla2017linear} employed a linear systems approach to derive geometric and photometric bounds of looking around corners with SPADs. 
Tsai et al.\ \cite{tsai2017geometry} improved on the technique proposed by Buttafava et al. using the arrival times of only first-photons coming from the hidden object.
Gariepy et al.\ \cite{gariepy2015detection} used SPADs to detect and track slow moving objects around the corners. 
Pediredla et al.\ \cite{pediredla2017reconstructing} designed a system based on SPADs to image hidden-rooms if one wall is visible from a window or door. 
Heide et al.\ \cite{heide2017robust} proposed algorithmic framework robust to partial occlusions in the hidden scene and recovered the hidden object with data captured from SPAD-based TCSPC system. 
All these methods are inherently affected by pile-up.  


\subsection{Pile-up compensation}
Most of the prior-art on pile-up compensation was targeted at lifetime applications and for the non-gated mode. Early work by Coates \cite{coates1968correction, coates1992analytical} proposed a pile-up algorithm to measure the radiative lifetimes. Davis and King \cite{davis1970single}  
proposed to use approximations for the Poisson process to develop a pile-up compensation technique. 
Luhmann \cite{luhmann1997statistics} created a calculation method to compensate for the dead-time in two-particle coincidence experiments.
Renfro et al.\ \cite{renfro1999pulse} proposed an iterative technique called saturate-and-compare for measuring fluorescence lifetime. 
Walker \cite{walker2002iterative} proposed an iterative technique to compensate for the pile-up under variable pulse energy and showed the results of simulations. 
Patting et al.\ \cite{patting2007dead} proposed a model-based approach to compensate for the dead time and the associated pile-up.
Rebafka et al.\ \cite{rebafka2011information} used concepts of information theory to decrease the acquisition time for lifetime imaging. 
Arlt et al.\ \cite{arlt2013study} studied the effect of the pile-up in integrated solid-state TCSPC sensors with small dead time. 

\begin{figure}
	\centering
	\includegraphics[width=5.250in]{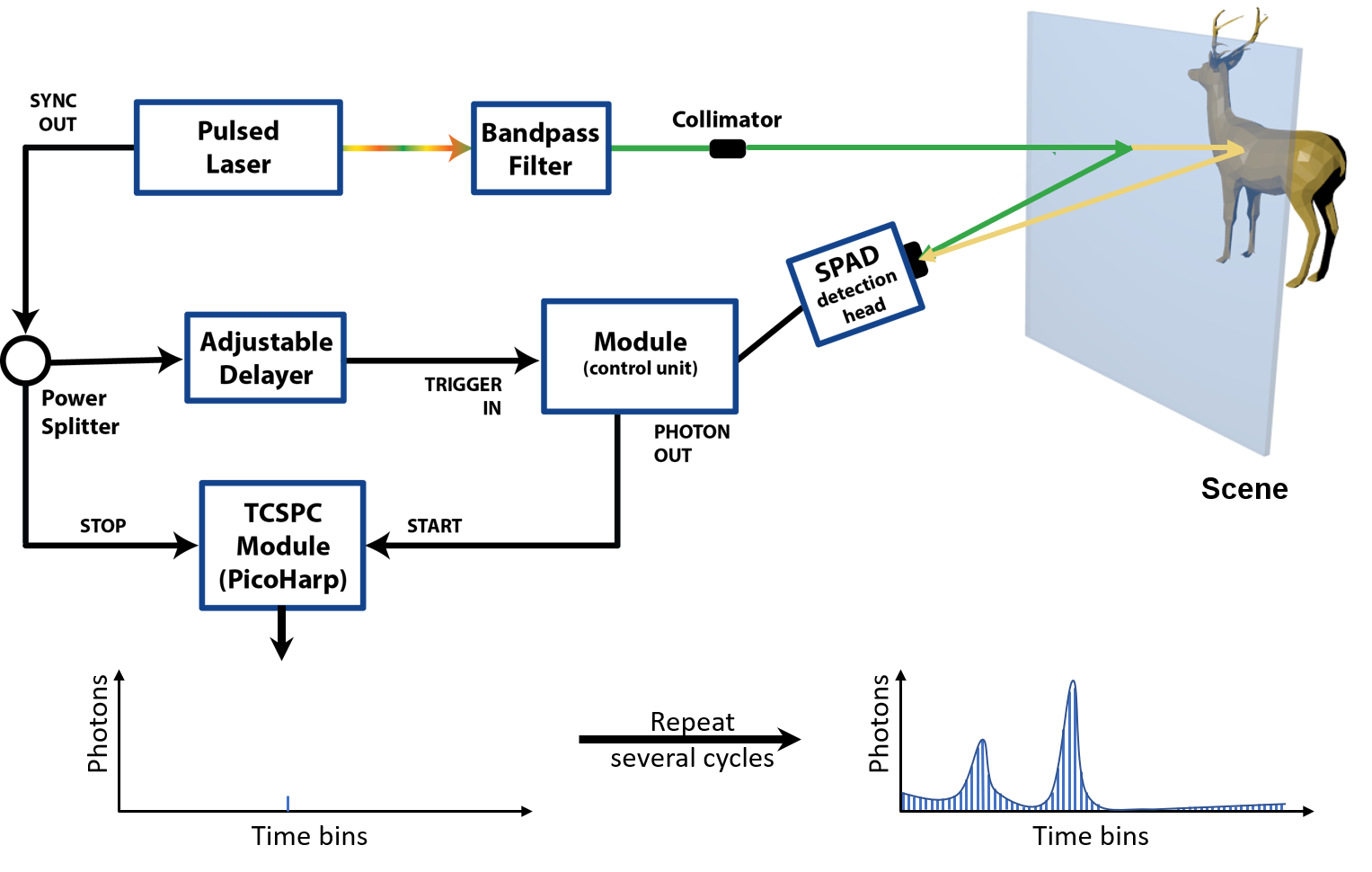} \\
	\caption{{\bf Operation of TCSPC:} Every time an optic pulse is fired by the laser, an electronic NIM (Nuclear Instrumentation Module standard) pulse is also sent to the TCSPC. The SPAD also sends a NIM pulse every time it detects a photon. TCSPC measures the time of travel of the optic pulse by computing the time difference of the NIM pulse sent by the SPAD with the nearest NIM pulse sent by the laser and increments the photon count of the appropriate time bin (that corresponds to photon travel time) by one. By counting a lot of photons received from the SPAD and appropriately binning them based on their arrival times, the TCSPC creates the transient response as shown in the figure.} 
	\label{Fig:TCSPCOperation}
\end{figure}
In all these techniques, the analysis is mostly experimental, and no guarantees on the efficacy of the recovered parameters are ever made, except maybe in \cite{rebafka2011information} where the goal is to minimize the acquisition time of time-resolved fluorescence for non-gated mode SPAD operation. These techniques do not apply mature signal processing approaches for pile-up compensation. In this work, we deploy some fundamental estimation-theoretic methods from signal processing literature on the transients captured by the SPAD-based TCSPCs. We believe that the vast literature of estimation theory \cite{kay1993statistical,scott2015multivariate} will further improve the results that we show in the paper and hence, we made both the code and data publicly available \cite{DropboxLink}.

\section{Modelling the transients captured by the SPAD-based TCSPC }
In this section, we will develop a probabilistic model for the histogram measured by the TCSPC. 
We will first explain the operation of the SPAD-based TCSPC following which we will develop a simulation framework and show how pile-up appears at high illumination power. 

\subsection{Operation of the SPAD-based TCSPC}

At a high-level, TCSPC works by binning the photons detected by the SPAD, based on their arrival times. The arrival time of a photon is calculated by measuring the time difference between the SPAD detection and the nearest illumination laser pulse. 
Figure~\ref{Fig:TCSPCOperation} shows a sketch of how TCSPC creates the transients.  

\begin{figure}
	\centering
	\includegraphics[width=3.800in]{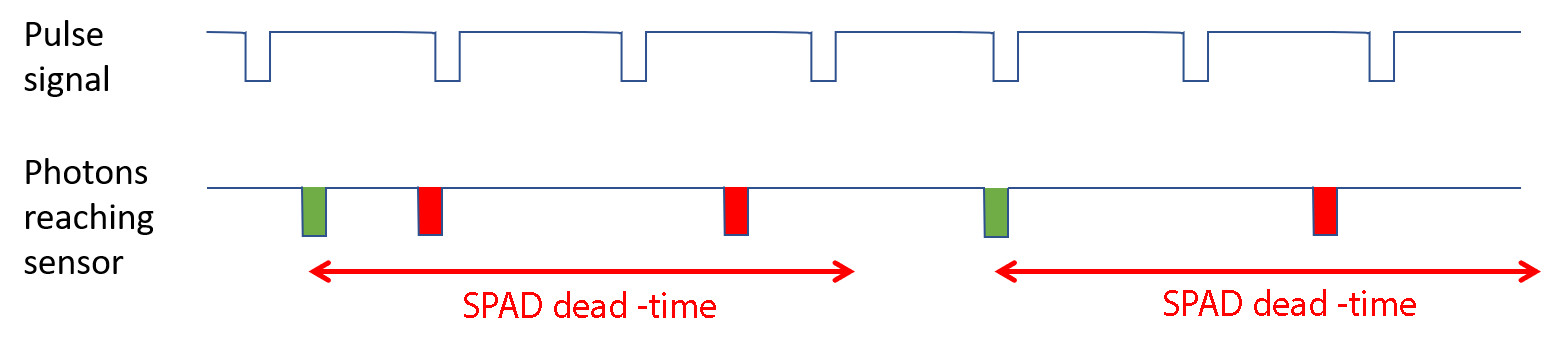} \\
	\caption{{\bf Photons detected in non-gated mode:} If a photon hits the active area of the SPAD during dead-time, it will not be detected and hence, the TCSPSC cannot measure it. We label the NIM pulse train (negative standard) sent by the laser as a pulse signal. The photons reaching the SPAD sensor are also shown in NIM standard where the green-colored NIM pulses (and the corresponding photons) are detected by the SPAD, and the red pulses are not detected.}
	\label{Fig:NonGatedModeOperation}
\end{figure}

When a photon hits the SPAD active area, an avalanche current builds up inside the device and is detected by the readout electronics (which generates the corresponding output signal). This current is then quenched by dedicated circuitry (quenching phase) in few nanoseconds and the detector is kept OFF for few tens of nanoseconds (hold-off time), in order to decrease its afterpulsing probability \cite{cova1996avalanche}. The sum of quenching time and hold-off time is commonly known as the SPAD dead-time. We illustrate the effect of dead-time in Figure~\ref{Fig:NonGatedModeOperation}. 
We notice that the photons corresponding to pulses that are colored red are not detected by the SPAD as they arrive during its dead-time. Note that the entire circuitry is operated with Nuclear Instrumentation Module (NIM) standard with negative true (-0.8 volts) and hence the pulses are loosely referred as NIM pulses throughout the paper. 
\begin{figure}
	\centering
	\includegraphics[width=5.250in]{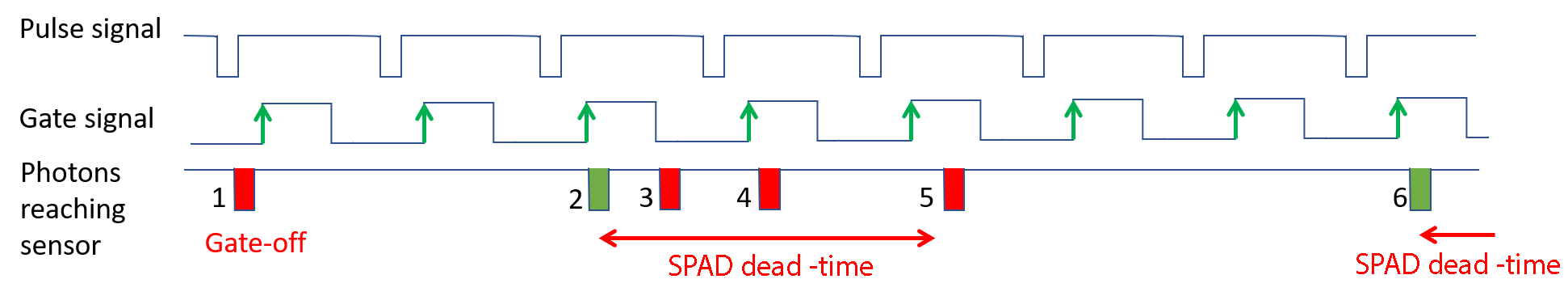} \\
	\caption{{\bf Photons detected in gated mode:} Gated mode SPADs accept photons only when gate signal is high (they turn ON at gate rising edge) and turn OFF when the signal goes low. After a photon detection and the related dead-time, the SPAD will turn back ON only at the rising edge of the next useful gate. In the above figure, the first photon is not detected as the SPAD is OFF when the photon arrived. Photons numbered 3 and 4 are rejected as they arrived during the dead-time. Photon 5 is a special case where the gate should be ON, but the SPAD is still OFF as there is no rising gate signal after the SPAD is recovered. Notice that, in every laser cycle, two photons cannot be detected by the SPAD, independent of whether the dead-time is more than the duration of the laser cycle or less. 
	} 
	\label{Fig:GatedModeOperation}
\end{figure}

Recently, gated mode operation for SPADs is becoming increasingly common. Gated mode operation helps SPADs to turn ON only when the photons are expected and hence, reject a lot of unwanted photons. In the gated operation mode, the SPAD turns ON only when the gate signal is high. The gate signal is synchronized with the laser pulse signal. Typically, the gate signal width is less than the duration between two consecutive laser pulses (if not, there is no real advantage in using the gate signal as the SPAD will be always ON). In gated mode operation, a photon is not detected if it arrives either when the gate signal is low, or during the SPAD dead-time. It should also be noted that, at the end of the dead-time period, the SPAD turns back ON only at the next useful rising edge of the gate signal. In Figure~\ref{Fig:GatedModeOperation}, the laser pulses and gate signals of a typical case are shown, along with the photons reaching the sensor. Among these photons, the ones tagged 1, 3, 4, 5 are not detected by the SPAD due to three different reasons. The first photon is rejected as the gate is OFF at its arrival time. The third and fourth photons are rejected as they arrive during the SPAD dead-time. The fifth photon arrives when the SPAD gate signal is high, but the SPAD is still OFF as the rising edge of the current gate precedes the end of the dead-time period, and this results in the rejection of the fifth photon as well (the SPAD will turn back ON at the next gate signal rising edge).

Notice that, in gated mode, the SPAD can only detect one photon per laser cycle. Even if the SPAD dead-time is shorter than the laser period the detector won't turn back ON in the same cycle, as the rising edge of the gate signal will only arrive in the next laser pulse. 
To minimize the photons that are not detected by the SPAD, it is recommended to keep the illumination intensity lower enough such that photon detection rate (number of photons detected per second) remains within 1-5\% of the laser repetition rate. In sharp contrast, the techniques developed in this paper enable the SPAD operation at photon detection rates close to 90\%.

\begin{figure}
	\centering
	$
	\begin{array}{cc}
	\includegraphics[width=2.60in]{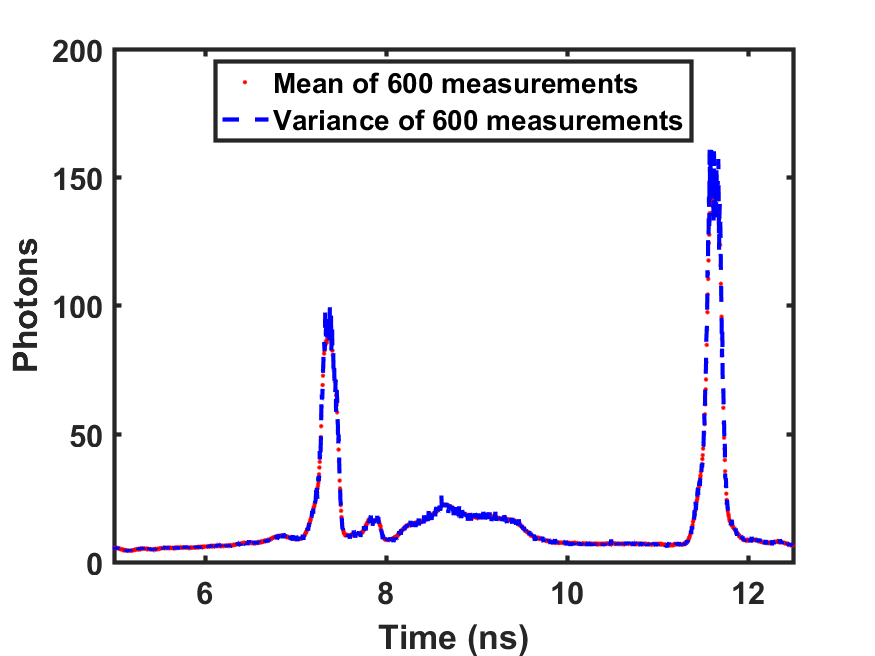} & 
	\includegraphics[width=2.60in]{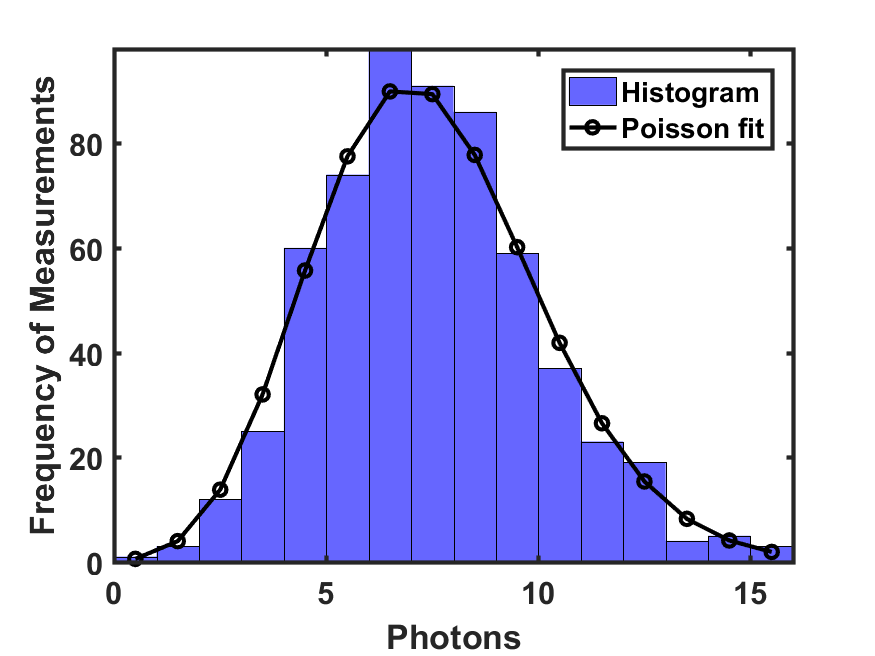} \\
	(a) & (b)
	\end{array}
	$
	\caption{{\bf Photons in a time bin are Poisson distributed:} We measured the transients at low illumination levels and background (no pile-up distortion) for 600 times with an integration period of 100 ms to measure the statistics of photons in a time bin. (a) shows the mean and variance of the transient measurement. We can observe that the mean and variance are nearly equal suggesting that the arrival process can be well modeled with the Poisson process. (b) shows the histogram of the frequency of measurements as a function of the number of photons received for a particular time bin. The histogram appears close to a Poisson probability mass function (PMF) suggesting that we can use the Poisson process to model the number of photons arriving in a time-bin.
	} 
	\label{Fig:PoissonAssumption}
\end{figure}

\subsection{Simulating SPADs}
\label{sec:SimulationSPADs}
Based on the understanding of the operation of the SPAD, we can simulate their operation if we know the photon arrival rate of each bin. 
In most literature, photon arrivals are modeled as the Poisson process. 
In this section, we empirically verify that this model is indeed accurate. In Figure~{\ref{Fig:PoissonAssumption}(a), we plot the mean and variance of the photons captured by our set up for 600 trials and integration period of 0.1 seconds per trial. The illumination is maintained at a very low value so that the pile-up does not affect the transient measurements. We observe that the mean and variance of any temporal bin are equal, just as it would if the arrival were a Poisson process. Further, the density of the photons in all temporal bin fits a Poisson density (Figure~{\ref{Fig:PoissonAssumption}(b)). Hence, we model the photon counts at any temporal bin as a Poisson distribution, given as 
		\begin{align}
		P(X_k=x) = \frac{\lambda_k^x e^{-\lambda_k}}{x!},
		\end{align}
		where $X_k$ is a random variable, denoting the number of photons arriving at the $k^{\textrm{th}}$ bin per laser pulse. $\lambda_k$ is the mean photon count of the $k^{\textrm{th}}$ bin.
		
		Next, we develop a simulation framework for the histogram measurements using the operating principles of the SPAD and the TCSPC coupled with the Poisson model for the photon count. The details are provided in Algorithm~\ref{ForwardModelAlgorithm}. The algorithm depends on the system parameters of the SPAD (namely integration time, dead time of the SPAD, laser repetition rate, and sampling rate of TCSPC), the illumination power, and the ideal transient response of the scene for one laser pulse ($\Lambda$). Notice that the algorithm is embarrassingly parallel and we have implemented it on a GPU, which is $\sim$1000$\times$ faster than a CPU implementation. The GPU version of the MATLAB code is available at \cite{DropboxLink}. Hernandez et al.\ \cite{hernandez2017computational} proposed a GPU accelerated simulation model for the SPAD operated in non-gated mode. Our algorithm is similar to that of Hernandez et al.'s algorithm, but for gated mode.
		
		\begin{algorithm}[!ttt]
			\caption{Simulation algorithm for the SPAD}\label{ForwardModelAlgorithm}
			\begin{algorithmic}[1]
				\State \textbf{Input:}
				\State $T \gets \text{Bin width of histogram (4~ps)}$
				\State $\mathcal{R} \gets \text{Repetition period of laser pulses}$
				\State $\Lambda \gets \text{Ideal transient (with integration time $\mathcal{R}$)} $; $\Lambda = \{\lambda_1, \lambda_2, \cdots, \lambda_N\}$.
				\State $N \gets \text{number of time-bins}$.
				\State $\mathcal{H} \gets \text{Dead time/Hold off time of the SPAD}$.
				\State $\tau \gets \text{Integration time}$
				\State \textbf{Initialization:}
				\State $h_k \gets 0;\,\,k=\{1, 2, \cdots N\}$
				\State $t \gets 0$
				\While{$t < \tau$}
				\State $\text{Photons}_i \gets Poisson(\lambda_i);\,\,i=\{1, 2, \cdots, N\}$
				\If{$\sum_i{\text{Photons}_i} > 0$}
				\State $k \gets \min\{i\,|\,\text{Photons}_i \ne 0\}$
				\State $h_k \gets h_k + 1$
				\State $t \gets t + \mathcal{H} + kT$
				\State $t \gets \mathcal{R} * \ceil{t/\mathcal{R}}$
				\Else
				\State $t \gets t + \mathcal{R}$
				\EndIf
				\EndWhile
				\State \textbf{Output:}  $h_k$
			\end{algorithmic}
		\end{algorithm}
		
		Using Algorithm \ref{ForwardModelAlgorithm}, we can observe the pile-up effect as we increase the illumination power. We consider the transient of a scene with two peaks and compute the histogram captured by the SPAD at higher illumination power. 
		In Figure~\ref{Fig:PileUp}(a), we show the normalized transient measured by the SPAD simulation along with the ground truth transient. We can notice that the photons appear to pile-up near the origin. 
		However, from Figure~\ref{Fig:PileUp}(b), we can observe that the photons are dropped at all the time bins; Photons from the time bins that are farthest from the origin are dropped more compared to the photons near the origin. 
		
		\subsection{Probabilistic forward model for TCSPC histogram}
		Let $\lambda_k, k \in \{1, \cdots, N\} $ be the ideal transient response per laser cycle at light level $I$. Note that $k$ denotes the bin index of the TCSPC histogram, starting from the bin immediately after the gate is turned ON. Hence, bin index `1' denotes the bin exactly after the gate is turned ON. The measurement by an ideal transient imager for a laser cycle will be $y_k = \mbox{Poisson}(\lambda_k)$, as noted in Section~\ref{sec:SimulationSPADs}. 
		
		%
		In each cycle of the laser, the SPAD can measure the arrival time of at most one photon. 
		%
		%
		Let $B_k$ denote of the event that the SPAD-based TCSPC measured the photon arrival in the $k^{\textrm{th}}$ bin.
		The event $B_k$ happens when no photons arrive in the first $k-1$ bins, and at least one photon arrives in the $k$-th bin, and therefore $B_k$ happens with the probability
		%
		$$P(B_k) = p_k = e^{-\sum_{i=1}^{k-1} \lambda_i} (1-e^{-\lambda_k}).$$ 
		We further define $B_0$ as a special event where no photons arrived at any time-bin in a laser cycle that SPAD is not under quenching mode. 
		Therefore,
		\[ P(B_0) = p_0 = e^{-\sum_k{\lambda_k}}. \]
		
		If all the arrival rates ($\lambda_k$s) are very small, then 
		\[ \forall k >0, \qquad P(B_k) \approx \lambda_k. \]
		Therefore at low photon count rates, the histogram captured by the SPAD is proportional to the transient response. 
		If the low photon count assumption is not valid, then the histogram is not proportional to the transient response. 
		We derive the histogram for all cases (both low illumination and high illumination) next.

		\begin{figure}
			\centering
			$
			\begin{array}{cc}
			\includegraphics[width=2.50in]{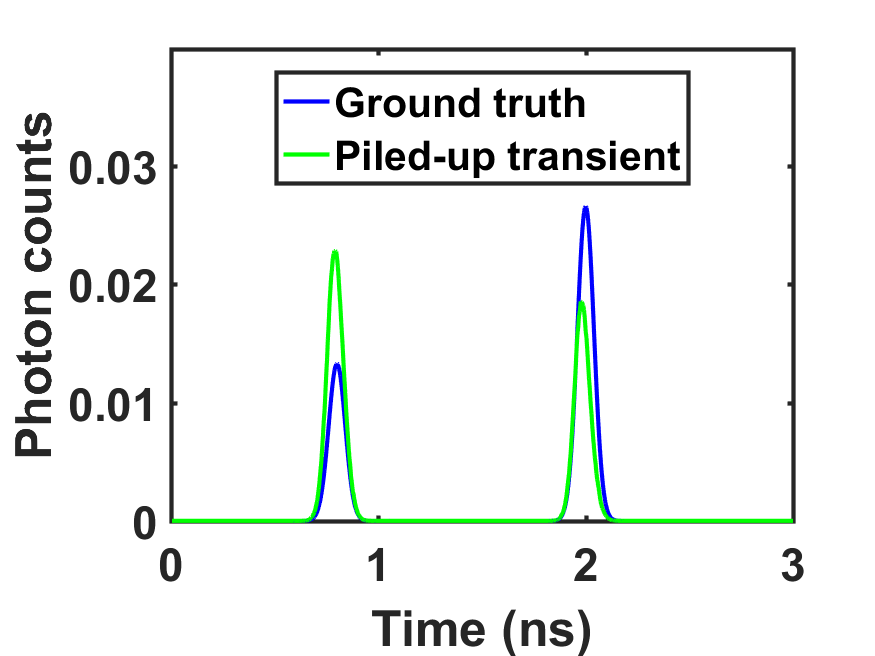} & 
			\includegraphics[width=2.50in]{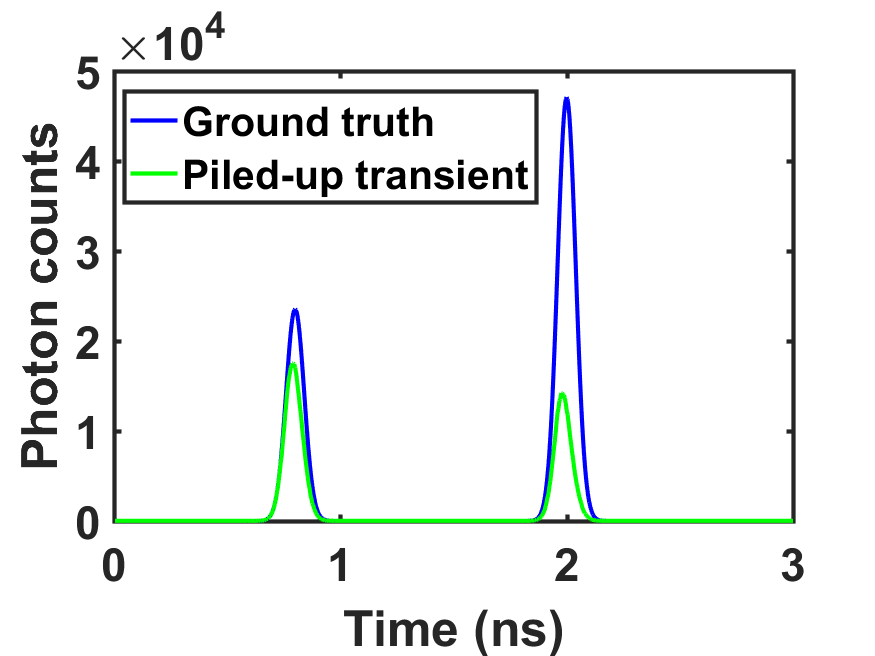} \\
			\mbox{(a) Normalized transients} & \mbox{(b) Un-normalized transients}
			\end{array}
			$
			\caption{{\bf Simulations showing the Pile-up distortion:} (a) When the illumination intensity is increased, the photon flux measured by SPAD-based TCSPSC system scales non-linearly with time of travel. Early transients tend to scale with intensity, and the later transients tend to decrease with the illumination power.  (b) However, photons arriving at all the time-bins are lost due to SPAD dead time. Photons at the earlier part of the transient have less probability to get dropped compared to the later part. 
			}
			\label{Fig:PileUp}
		\end{figure}
		Let $H = \{h_0, h_1, h_2,  \cdots, h_n\}$ be an observed histogram for integration time $\tau$; note that we have introduced the count of no photons arrivals ($h_0$) in the observed ($H$) even though we do not directly observe $h_0$. We will compute $h_0$ by first computing the number of cycles the SPAD is active ($s$) and then subtracting the total number of cycles ($\sum_{k=1}^n h_k$) we have a photon arrival. 
		
		The total number of laser cycles in integration time $\tau$ is $\frac{\tau}{R}$ where $R$ is the repetition time of the laser.
		Recall from Figure~\ref{Fig:GatedModeOperation} that the number of cycles $s$ that the SPAD is active  is less than or equal to the number of laser cycles.
		Each time the SPAD detects a photon in $k^{\textrm{th}}$ time bin, the SPAD will not detect the photon for  $\frac{\mathcal{H}+kT}{R}$ laser cycles, where  $\mathcal{H}$ is hold off or dead time and $T$ is bin width of histogram.
		Therefore the total number of laser cycles the SPAD is active is given by 
		\begin{align}
		s = \frac{\tau}{R} - \left( \left\lceil{\frac{\mathcal{H}+kT}{R}} \right\rceil - 1\right)  \sum_{k = 1}^n h_k.
		\end{align}
		
		If the hold-off time ($\mathcal{H}$) is very small compared to repetition time of the laser ($R$) $\mathcal{H} \ll R$ then $s = \frac{\tau}{R}$, i.e., the SPAD is active for all the laser cycles. Note that the above equation is valid only for gated SPAD operation as non-gated operation can measure more than a photon per laser pulse. 
		
		We now have all the ingredients to compute the probabilistic forward model for the TCSPC histogram. 
		Our first observation is that the histogram measured by the SPAD can also be modeled as a multinomial distribution. 
		Recall that the multinomial distribution models the event of tossing a $p$-faced dice $n$ times and observing the vector $(x_1, x_2, \cdots x_p)$, where $x_k$ denotes the number of times $k^{\textrm{th}}$ face showed up in the toss.
		We can now map each event in $\{ B_0, B_1, \ldots, B_N\}$ to a face of the dice; hence $p = N+1$. 
		The number of tosses $n$ is equal to the number of the cycles $s$ that the SPAD is active. 
		Hence, the probability of observing the histogram $H$  is given as
		\begin{align}
		P(H|\Lambda) &=  \frac{s!}{h_0! h_1! \ldots h_N!} p_0^{h_0} p_1^{h_1} \ldots p_N^{h_N}  \nonumber \\
		&= \frac{s! }{(s-\sum_{k=1}^n h_k)! h_1! \ldots h_N!} e^{-(s-\sum_{k=1}^n h_k) \sum_{k=1}^n{\lambda_k}} \Pi_{k=1}^N  e^{-h_k \sum_{j=1}^{k-1} \lambda_j} (1-e^{-\lambda_k})^{h_k} \nonumber  \\
		&= \Pi_{k=1}^{N} \binom{s-\sum_{j=1}^{k-1}h_j}{h_k} 
		\left(1-e^{-\lambda_k}\right)^{h_k}\left(e^{-\lambda_k}\right)^{s-\sum_{j=1}^{k}h_j} \nonumber \\
		&= \Pi_{k=1}^{N} P(h_k| h_1, h_2, \cdots h_{k-1}, \lambda_k).
		\label{Eq:PDF}
		\end{align}
		
		Equation (\ref{Eq:PDF}) is the probabilistic forward model that gives the relationship between the histogram measured by TCSPC and the true photon arrival rates. 
		Equation (\ref{Eq:PDF}) states that the probability mass function (pmf) of the histogram of the $k^{th}$ bin conditioned on the histogram observed in all the previous time bins is binomial distributed and the true photon arrival rate of the $k^{th}$ bin. 
		In the next sections, we will derive inverse models to computationally recover the underlying transients given the histogram measured by TCSPC.

		\begin{figure}
			\centering
			\includegraphics[width=5.25in]{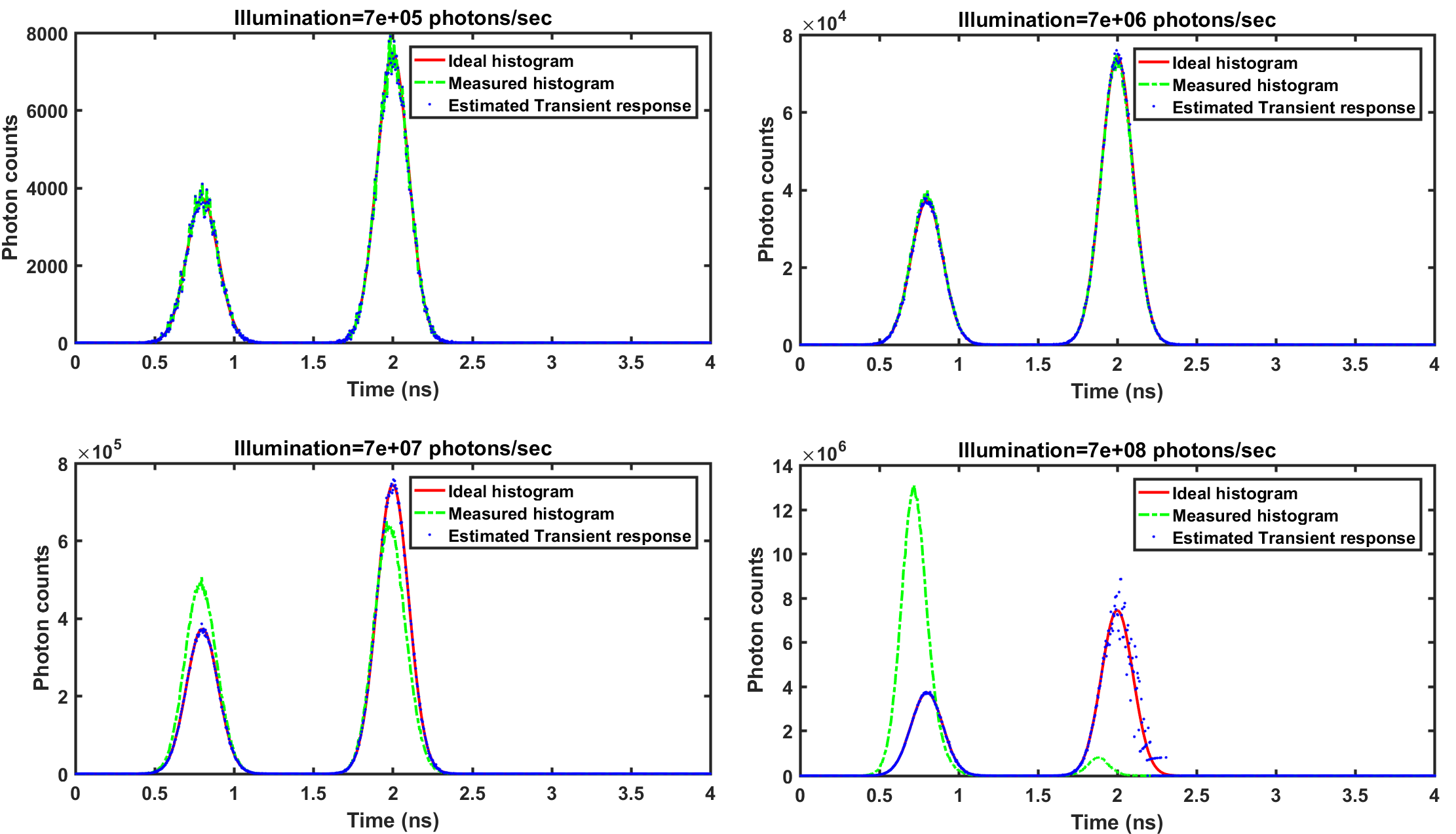} 
			\caption{{\bf Simulation of ML performance for transient recovery:} 
				We plot the ideal histogram, measured histogram, and estimated transient response from ML estimate for various illumination intensities. 
				We can observe that even though the measured histogram from the SPAD is quite different from the ideal transient response, our estimation algorithm can recover the ideal transient response. 
				As illumination power increases, noise in later time bins is relatively higher than the early bins; The number of photons (hence signal) in the early bins is higher than the later bins causing less noise in early bins.
			}
			\label{Fig:MLEstimate}
		\end{figure}
		
		\section{Compensating pile-up} 
		Given the forward model and the experimental photon count measurements in the form of a histogram, we now seek to estimate the transient response $\Lambda$. We derive two estimates for the transient response: first, a maximum-likelihood (ML) estimate and, second, a conjugate prior-based maximum-a-posteriori (MAP) estimate.
		
		\subsection{ML Estimate}
		\label{ssec:MLEstimate}
		Given an observation $H = \{h_0, h_1, h_2,  \cdots, h_n\}$, the likelihood is simply the conditional probability $$P(H  | \lambda_1, \cdots \lambda_n).$$
		Maximizing the log-likelihood with respect to $\lambda$s provide the ML-estimate of $\lambda_k$: 
		\begin{align}
		\displaystyle \widehat{\lambda}_k = \log \left(1+\frac{h_k}{s-\sum_{i=1}^{k}h_i}\right).
		\end{align}
		
		In Figure~\ref{Fig:MLEstimate}, we show the efficacy of our estimate in recovering the underlying transient signal from the piled-up transient responses shown in Figure~\ref{Fig:PileUp}.
		
		\subsubsection{Cramer-Rao bound}
		
		The Cramer-Rao bound provides a lower bound on the error of unbiased estimators. In this section, we will derive the Cramer-Rao lower bound on the estimation of transients, explain the implications of the bound, and show that ML-estimation reaches this bound for practical scenarios. 
		
		The Cramer-Rao bound for the variance of any unbiased estimator of the transient rate $\lambda_k$ is equal to
		\begin{align}
		\frac{1-e^{-\lambda_k}}{s \displaystyle e^{-\sum_{j=1}^{k}\lambda_j}}.
		\label{Eq:CRBound}
		\end{align}
		We provide the derivation in the Appendix.
		Since $\lambda_k$ is directly proportional to the illumination intensity, increasing illumination increases the error in the ML estimate. 
		Therefore, for very large illumination, it will become increasingly difficult to capture the transient response with SPADs, which is intuitive as well -- for very large illumination power, the first arrival photon at cycle will occur at first time bin leading to a massive information loss. 
		
		A second observation from (\ref{Eq:CRBound}) is that increasing integration time decreases the estimation error. As integration time increases, the number of cycles $s$ will increase, decreasing the estimation error. Therefore, for any finite illumination power for a given system, the estimation error can be made arbitrarily small by increasing integration period. 
		
		A third observation from (\ref{Eq:CRBound}) is that bins with higher true transient value tend to have a higher error. In fact, bins that have zero count rate will have zero estimation error. As the arrival process is Poisson, higher the mean, higher will be the variance leading to this condition.
		
		A fourth observation from (\ref{Eq:CRBound}) is that early histogram bins tend to have smaller estimation error than later histogram bins. For two time bins with same $\lambda$, the denominator of (\ref{Eq:CRBound}) will be smaller for the later histogram bin. Hence, the error of the later histogram bin will be higher. This is also intuitive from the pile-up effect that causes more photons in the early time bins, hence less variance, compared to the later time-bins.

		In Figure~\ref{Fig:MAPvsMLEstimate}, we show the Cramer-Rao lower bound on the estimation error and the error of the ML estimate for a fixed integration time of 10~$\mu$s. We can observe that they are almost equal implying that the error in ML estimate converges to Cramer-Rao bound. This is a well-known phenomenon in signal processing community and happens because ML estimate is asymptotically efficient \cite{ross2014introduction} and reaches Cramer-Rao bound for a large number of cycles.
		As the repetition rate of the laser is in the order of few million cycles per second, the ML estimate reaches Cramer-Rao bound even for exposure duration as small as 10~$\mu$s. Hence, for most practical SPAD usages, the ML-estimate is the best unbiased estimate for solving pile-up.
		
		\begin{figure}
			\centering
			$
			\begin{array}{c}
			\includegraphics[width=5.25in]{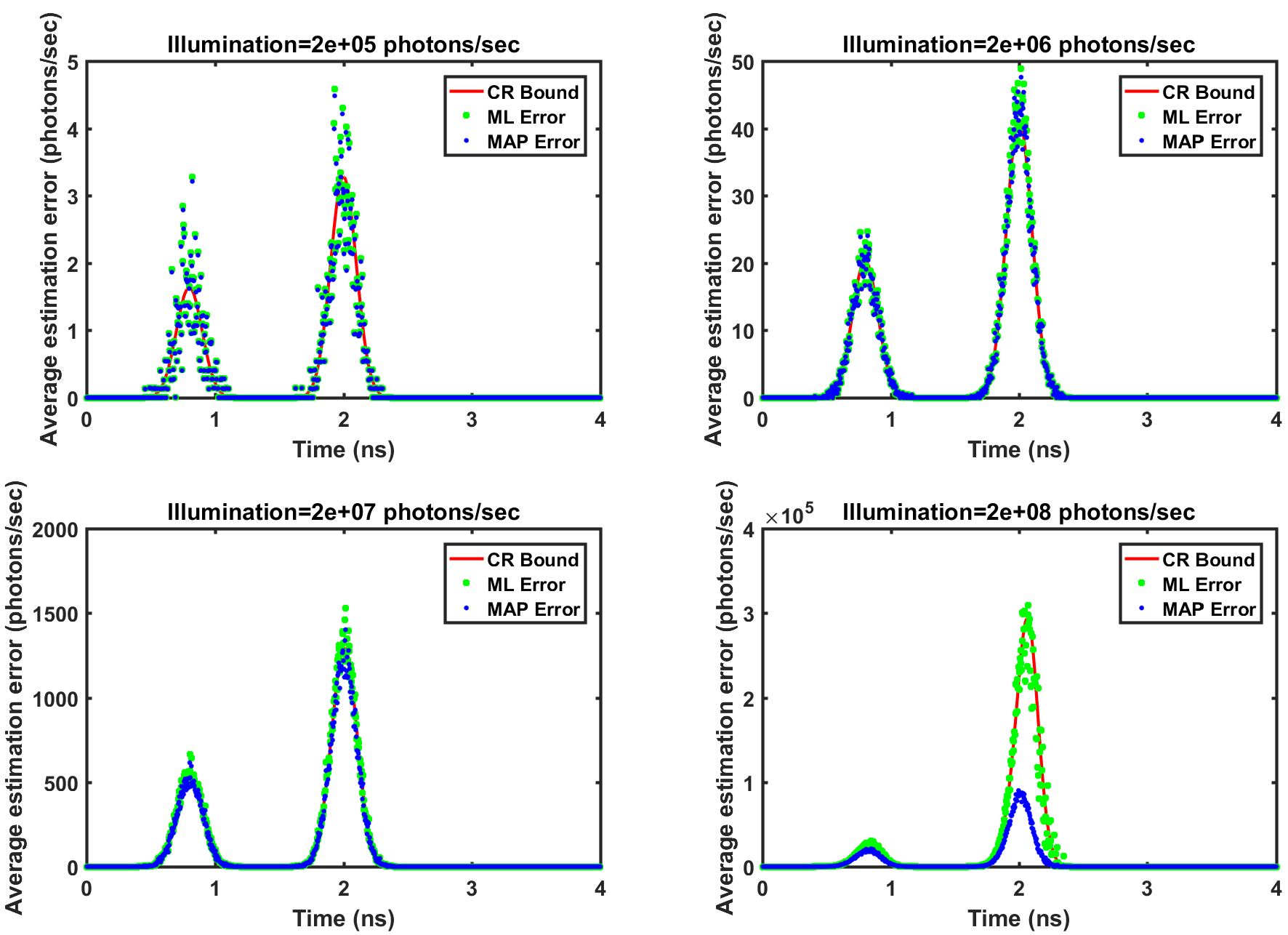} \\
			\end{array}
			$
			\caption{{\bf Simulation results showing that estimates with prior knowledge (MAP) can perform better:} Incorporating prior knowledge (Bayesian) of the transient response can result in lower estimation error, especially at high illumination intensity. We used the conjugate prior for multinomial distribution, which is a beta-distribution, as the prior so that the resulting estimate has a closed form solution. From the plots, we can notice that the MAP error is smaller than the ML error, and even CR-bound because of the prior knowledge. Note that MAP estimator is biased and also assumes that $\Lambda$ is not deterministic and hence is able to beat the Cramer-Rao bound.
			}
			\label{Fig:MAPvsMLEstimate}
		\end{figure}
		\subsection{MAP Estimate}
		\label{ssec:MAPEstimate}
		One way to circumvent the limits imposed by Cramer-Rao bound (especially at smaller $s$) is to use prior knowledge on $\Lambda$ to get a better estimate. 
		This can be achieved by using the MAP estimate with various priors on $\Lambda$. We can assume that $\Lambda$ is a sparse vector and use $\ell_1$ regularizer, which is equivalent to modeling $\lambda_k$s as being Laplacian distributed. 
		We can also use the Gaussian distribution ($\ell_2$ regularizer) on $\Lambda$, which is another standard prior. 
		%
		However, both Laplacian and Gaussian priors can result in negative values for $\Lambda$ with non-zero probability. Also, Laplacian and Gaussian priors do not give closed form solutions for $\Lambda$ and iterative techniques must be employed for solving them. Hence, we use the conjugate prior for the multinomial distribution, which guarantees a closed form solution for $\Lambda$.
		
		The conjugate prior for the binomial distribution is the beta distribution. Since $P(H|\Lambda)$ in (\ref{Eq:PDF}) is a multinomial distribution ($n$-dimensional binomial), we employ $n$-dimensional beta distribution as its conjugate prior. The beta distribution prior restricts the values of all $\lambda_i$s to the positive side of the real line.
		Hence, the optimization problem that we have is 
		\begin{align}
		\widehat{\Lambda} = \min_{\Lambda} -\log P(H | \Lambda) - \log \left( \Pi_{k=1}^N \frac{(e^{-\lambda_k})^{\alpha_k - 1}(1-e^{-\lambda_k})^{\beta_k - 1}}{B(\alpha_k, \beta_k)} \right), \nonumber
		\end{align}
		where $B$ is beta function and $\alpha_k$ and $\beta_k$ are shape parameters. 
		Solving the above equation, we get the estimate $\displaystyle \widehat{\lambda}_k = \log \left(1+\frac{h_k + \alpha_k - 1}{s-\sum_{i=1}^{n}h_k + \beta_k - 1}\right)$. Note that when $\alpha_k = \beta_k = 1$, all values of $e^{-\lambda_k}$ are equally likely, and we get ML estimate. Also, for a large number of cycles s, the data term will dominate the prior knowledge term and will result in ML estimate. 
		
		We know that many $\lambda_k$s are close to zero as the transient response is typically sparse. Therefore, we choose the prior to be more skewed to increase the probability for $\lambda_k=0$ by choosing $\alpha_k=1$ and $\beta_k =10$.
		
		In Figure~\ref{Fig:MAPvsMLEstimate}, we show that at low integration times ($\sim 10\ \mu$s), the MAP estimate can result in an error that is smaller than both the ML estimate and the CR bound (lower bound on all unbiased estimates). However, as the integration time increases ($\sim 100\ \mu$s), this improvement became very insignificant even under high illumination conditions. The hardware setup in our experiments cannot measure with integration times less than a millisecond, and hence, the comparison between ML estimate and MAP estimate is limited to theoretical and simulation analysis in this paper.

		%
		%
		%
		
		
		\section{Experimental setup}
		\label{sec:SetUp}
		The experimental setup developed for the validation of the pile-up compensation framework is shown in Figure~\ref{Fig:ExpSetup}. A supercontinuum laser source (Super-K extreme, NKT photonics A/S) produces light pulses with 5~ps width at 78 MHz of repetition rate, in a wavelength range from 400 nm to 2400 nm. The repetition rate is controlled with a pulse-picker to decrease the repetition rate up to 2.1 MHz. A tunable bandpass filter (Super-K varia, NKT photonics A/S) is used to select the operating wavelength interval of 550 nm $\pm$ 20 nm, achieving an average light power of around 0.25 mW. The fiber-coupled output of the bandpass filter is collimated (PAF-X-11, Thorlabs Inc.) and passed through a double polarizer to adjust the illuminating power of the entire scene. The beam is then divided into two paths using a 50/50 beam splitter (CM1-BS013, Thorlabs Inc.) and directed toward two white cardboard-made patches, placed at different distances from the beam splitter. A part of the photons scattered from the patches can reach the detection system at different time instants, giving rise to two well-separated peaks in the TCSPC histogram. 
		
		\begin{figure*}
			\centering
			\includegraphics[width=5.25in]{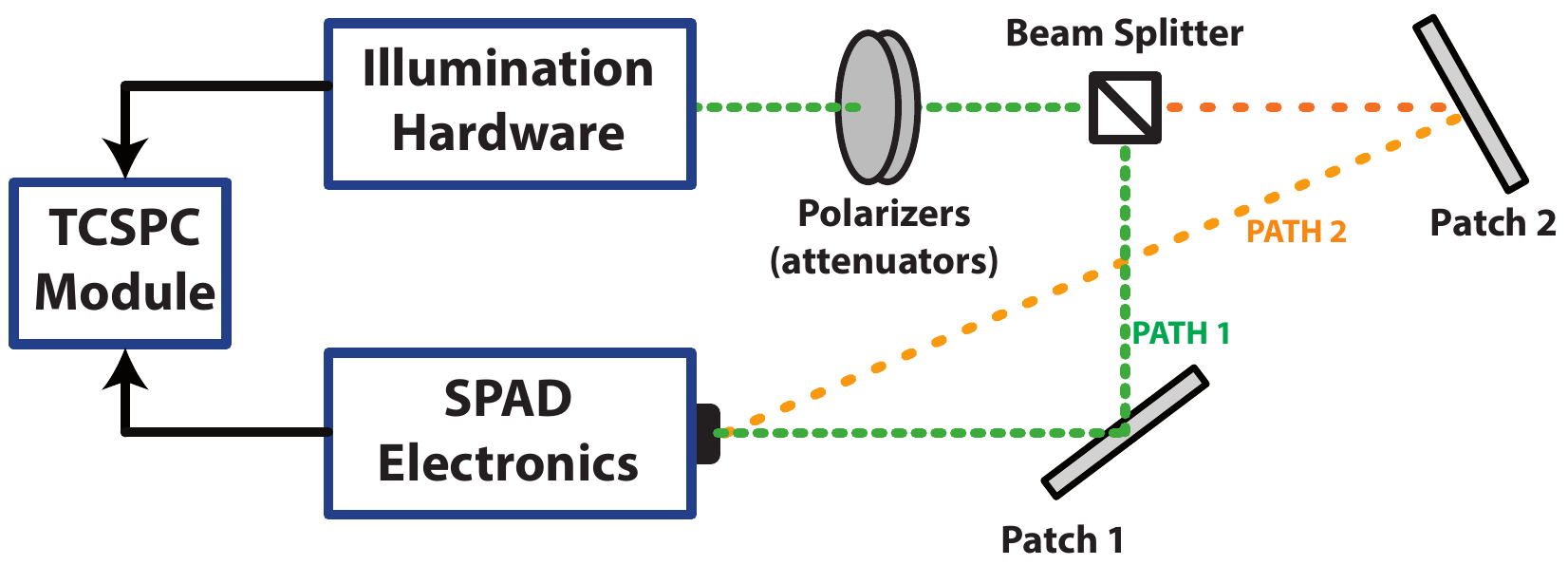} \\
			\caption{{\bf Experimental setup:} The set up consists of the SPAD based TCSPC system described in Figure~\ref{Fig:TCSPCOperation}. We used a single-pixel SPAD as the goal is to demonstrate the pile-up and the compensation algorithm. The illumination power is controlled with a polarizer pair. A beam splitter is used to create two paths of different path length. 
			}
			\label{Fig:ExpSetup}
		\end{figure*}
		The detection system is composed of a fast-gated single-photon counting module \cite{buttafava2014time}, hosting a single-pixel 50~$\mu$m active-area diameter CMOS SPAD. The module exhibits a photon detection efficiency higher than 30\% at 550 nm, with a temporal resolution of 35 ps FWHM (full-width at half maximum) and a dark-count noise lower than 100 counts per second (cps). The SPAD can be operated either in gated-mode (with adjustable gate width and less than 200 ps transition times) or in free-running mode, both with selectable hold-off time (120 ns in these experiments, determining a maximum count rate of 7.77 Mcps). The photon-out output pulse from the detection module acts as a START for the time-measurement system, while the synchronization signal from the laser acts as STOP. As a time-measurement system, we used a PicoHarp 300 TCSPC module (PicoQuant GmbH), able to acquire photon arrival times with 4 ps resolution, 260 ns of range and 10 Mconv/s of maximum conversion rate. The TCSPC histograms are downloaded to a PC via a USB 2.0 interface for data analysis. When operated in gated-mode, the SPAD module also needs a trigger signal for generating the ON-time window; this signal is derived from the laser synchronization signal and properly delayed (using an adjustable picosecond, Micro-photon Devices s.r.l.), being able to time-shift the ON-time window position of the SPAD respect to the laser pulse.  
		
		\begin{figure*}
			\centering
			\includegraphics[width=5.3in]{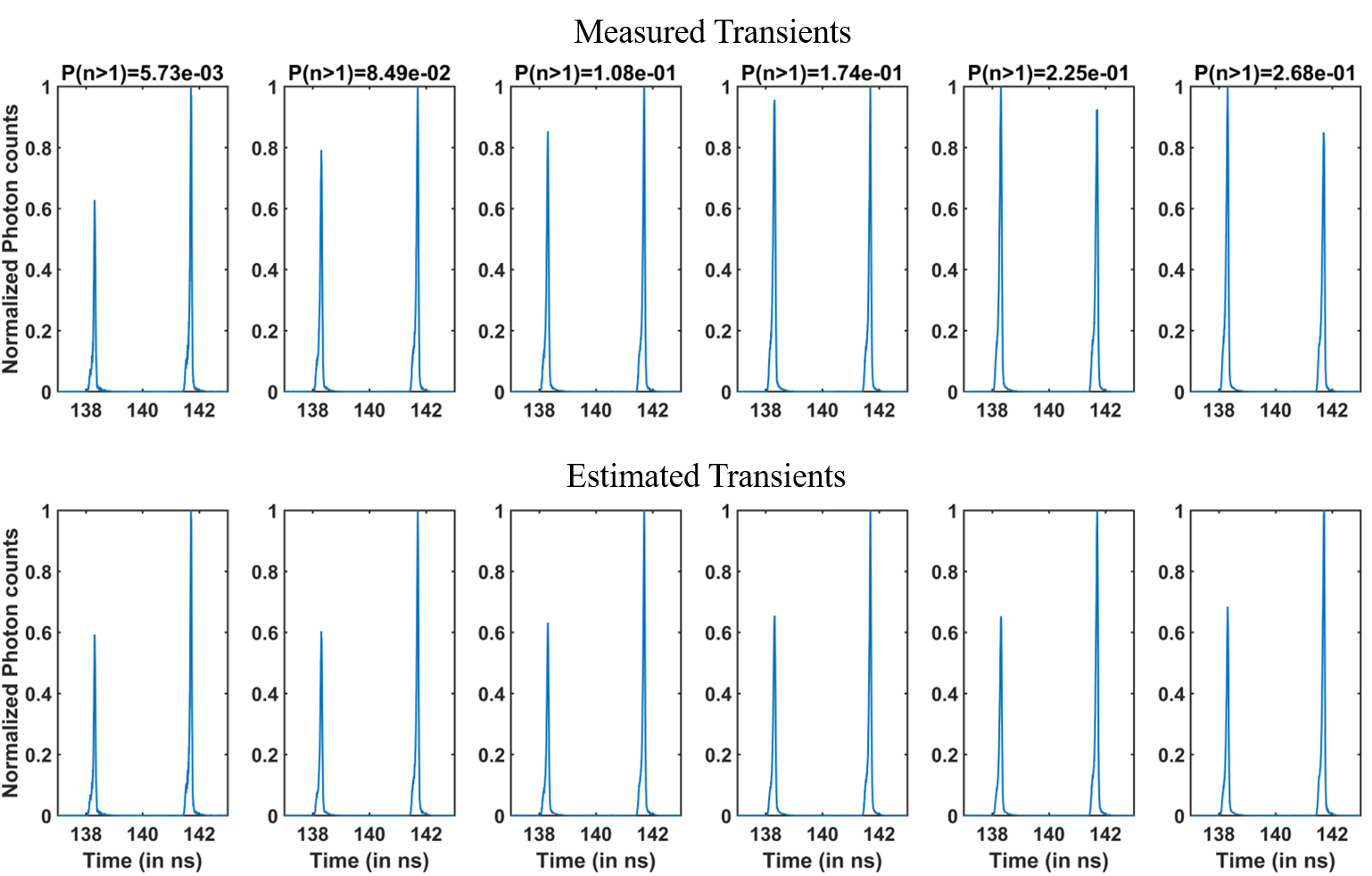} \\
			\caption{{\bf Experimental demonstration of compensating pile-up caused by strong illumination:} By aligning the orientation of the polarizers, the illumination intensity is increased. The probability of two photons arriving at the SPAD in a given laser cycle increases with increased intensity causing the pile-up. In the top row, as the illumination is increased (from left to right), the first peak appears to be stronger than the second peak due to pile-up effect. In the bottom row, we can notice that our pile-up compensation algorithm scales the transients appropriately to compensate for the pile-up.
			}
			\label{Fig:IlluminationCharacterization}
		\end{figure*}
		\section{Experiments}
		%
		
		We performed our experiments in the  two-path scene shown in Figure~\ref{Fig:ExpSetup}. If the illumination intensity is high-enough or if the background photons are strong enough then multiple photons will arrive at the SPAD with-in the same laser cycle. 
		
		Figure~\ref{Fig:IlluminationCharacterization} shows the effect of increasing the illumination power for the two-path scene. As the illumination is increased, the first peak becomes stronger, and the second peak decreases. ML estimate from the proposed framework recovers the relative amplitude of both the peaks accurately. We quantify the intensity of the illumination with $p(n>1)$ metric which represents the probability of two or more photons arriving at the SPAD in a laser cycle. From (\ref{Eq:PDF}), this value is given by $1-(1+\sum\lambda_k)\exp(-\sum\lambda_k)$. However, as we do not have access to $\lambda_k$ in real data, we approximate $\lambda_k$ with $\widehat{\lambda}_k$.
		
		Figure~\ref{Fig:BackgroundCharacterization} shows the effect of increasing the background for the two-path scene.  When the background (ambient illumination) is increased steadily, the first peak starts appearing stronger than the second peak, and a further increase in background intensity resulted in transients where the peaks due to the objects are indistinguishable from the peaks due to noise at the early part of the transient. Note that the gate is turned ON around 144~ns for a duration of 5~ns. We can observe that our framework is able to recover the transient even when 90\% of the photons are not detected by the SPAD-based TCSPC technique due to the dead time.
		
		\begin{figure*}
			\centering
			\includegraphics[width=5.3in]{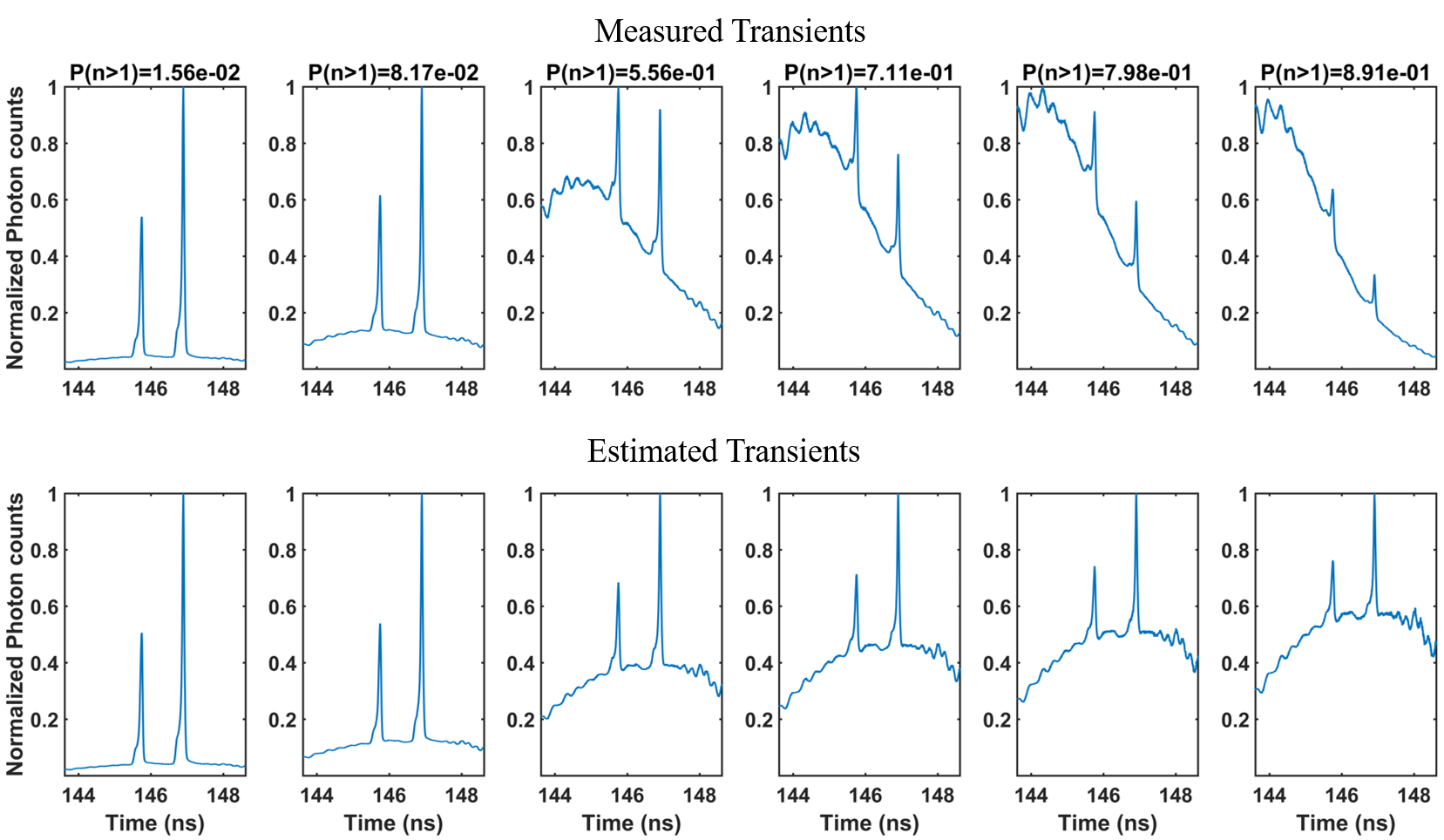} \\
			\caption{{\bf Experimental demonstration of compensating the pile-up caused by strong background (ambient illumination):} As the background illumination increases, the probability of more than one photon arriving at the SPAD per laser cycle also increases causing the pile-up, as shown in the figure. When less than 10\% of the cycles lose photons, the transients measured by the SPAD are not corrupted. However, as background is increased, the transients are heavily corrupted to a point that we cannot distinguish transients from noise. Our algorithm could recover the transients even when 90\% of the photons are dropped by the SPAD. 
			}
			\label{Fig:BackgroundCharacterization}
		\end{figure*}

		%
		%
		

\section{Conclusions}
To summarize, when multiple photons arrive at the SPAD in less than a laser cycle, only first photon is detected, and other photons are dropped by the SPAD. We model the photons accepted by the SPAD as a function the true-photons arriving at the SPAD. By inverting this model, we recover the lost photons computationally. 

Our inversion algorithm allows for high-illumination operation of the SPAD decreasing the integration time. 
SPAD-based techniques such as transient imaging or looking around the corners currently have poor integration times (order of minutes) and cannot image dynamic phenomenon. 
The integration times of these imaging methods can potentially be decreased by first employing a high power laser and then compensating for the pile-up applying the techniques in this paper. 

Data-driven priors, based on the imaging application can further improve the transient estimation similar to what we have shown in Section~\ref{ssec:MAPEstimate}. However, usage of data-driven priors mostly will not lead to simple $\mathcal{O}(n)$ algorithms that we have focused in this paper, but may potentially decrease the integration times further leading to real-time operation of the non-line-of-sight imaging or tracking around the corners. 


\section*{Acknowledgments}
This work was supported in part by NSF CAREER Grants IIS-1652633 and CCF-1652569, DARPA REVEAL grants HR0011-16-C-0028 and HR0011-16-2-0021, ONR grant N00014-15-1-2735, and the Big-Data Private-Cloud Research Cyberinfrastructure MRI-award funded by NSF under grant CNS-1338099.

\section*{Appendix}
\subsection*{Proof for CR-Bound}
Fisher information matrix is $\displaystyle I_{m,k} = - E\left[\frac{\partial^2}{\partial \lambda_m \partial \lambda_l} \log(P(H;\Lambda))\right]$

From (\ref{Eq:PDF}), due to separability of $\lambda_m, \lambda_l$, $I_{m,k} =0\,\forall\,m \ne k$. If $m=k$, we have 
\begin{align}
I_{m,m} &= -E\left[\frac{\partial^2}{\partial^2 \lambda_m} \log(P(H;\Lambda))\right]  \nonumber \\
&= -E\left[\frac{\partial}{\partial \lambda_m} \left(1+\frac{h_n}{1-e^{-\lambda_m}}e^{-\lambda_m}\right)\right] \nonumber \\
&= \frac{e^{-\lambda_m}}{(1-e^{-\lambda_m})^2} E[h_n]\nonumber \\
&= s\frac{e^{-\sum_{i=1}^{m}\lambda_m}}{1-e^{-\lambda_m}} \nonumber 
\end{align}
The Cramer-Rao bound states that cov$_\Lambda \ge I(\Lambda)^{-1}$. As $I(\Lambda)$ is a diagonal matrix, we have 
\begin{align}
Var(\widehat{\lambda}_m) &\ge \frac{1}{s}\frac{1-e^{-\lambda_m}}{e^{-\sum_{i=1}^{m}\lambda_m}} \nonumber
\end{align}

\end{document}